\documentclass[prb,twocolumn,showpacs,preprintnumbers,amsmath,amssymb]{revtex4}
\usepackage{amsfonts}
\usepackage{latexsym}

\begin{document}

\title{
Theory of terahertz electric oscillations by supercooled \\
superconductors}

\author{Todor M. Mishonov}\email[E-mail: ]{Mishonov@phys.uni-sofia.bg}
\author{Mihail T. Mishonov}

\affiliation{Department of Theoretical Physics,
Faculty of Physics, University of Sofia St. Kliment Ohridski,\\
5 J. Bourchier Blvd., 1164 Sofia, Bulgaria}

\begin{abstract}
We predict that below $T_c$ a regime of negative differential
conductivity (NDC) can be reached. The superconductor should be
supercooled to $T<T_c$ in the normal phase under DC voltage. In
such a nonequilibrium situation the NDC of the superconductor is
created by the excess conductivity of the fluctuation Cooper
pairs. We propose NDC of supercooled superconductors to be used as
an active medium for generation of electric oscillations. Such
generators can be used in the superconducting electronics as a new
type THz source of radiation. Oscillations can be modulated by the
change of the bias voltage, electrostatic doping by a gate
electrode when the superconductor is the channel of a field effect
transistor, or by light. When small amplitude oscillations are
stabilized near the critical temperature $T_c$ the generator can
be used as a bolometer. The essential for the applications NDC is
predicted by the solution of the Boltzmann kinetic equation for
the metastable in the normal phase Cooper pairs. Boltzmann
equation for fluctuation Cooper pairs is a result of
state-of-the-art application of the microscopic theory of
superconductivity. Our theoretical conclusions are based on some
approximations like time dependent Ginzburg-Landau theory, but
nevertheless can reliably predict appearance of NDC. NDC is the
main ingredient of the proposed technical applications. The
maximal frequency at which superconductors can operate as
generators is determined by the critical temperature $\hbar
\omega_{\rm{max}} \sim k_B T_c.$ For high-$T_c$ superconductors
this maximal frequency falls well inside the terahertz range.
Technical conditions to avoid nucleation of the superconducting
phase are briefly discussed. We suggest that nanostructured
high-$T_c$ superconductors patterned in a single chip can give the
best technical performance of the proposed oscillator.
\end{abstract}


\maketitle

\section{Introduction}
%
Due to their low ohmic dissipation superconductors find a lot of
technical applications. Superconductors can be used in resonators,
cables, electromagnets, transformers, electrical engines and
generators. As another important technical application we wish to
mention the SQIDs based on Josephson effect. However, up to now
there is a little progress in the use of superconductors as active
elements in electronic circuits such as amplifiers and generators
integrated in superconducting electronics. The present work falls
in this still uncompleted technical niche. We are suggesting how
superconductors, and especially high-$T_c$ superconductors, can be
used as generators of high frequency electric oscillations
including the THz range. When the superconductor is implemented as
a thin layer the work of the generator can be influenced by the
electrostatic charge modulation as in a gate transistor, by heat,
light or simply by the change of the DC bias voltage. Such a way
oscillations can be modulated, the generator can be used as an
optoelectronic device or as a bolometer. The electric oscillations
are generated by negative differential conductivity (NDC) of a
superconductor in non equilibrium condition. The superconductor is
supercooled in the normal phase below the critical temperature
$T_c$ under a constant electric field. The electric field prevents
transition of the superconductor to the superconducting phase
state. NDC is coming from so called fluctuation conductivity which
is strongly expressed for cuprate superconductors having smaller
coherence length.

\section{Physical model}
%
\subsection{Qualitative consideration and analogies}
%
First we will describe the idea how the negative differential
conductivity (NDC) can be created in a superconductor. Let us
consider the physical processes which happen in a supercooled in
the normal state superconductor under voltage. The material is in
a normal state and in his volume thermally activated Cooper pairs
are continuously created. This stochastic process is analogous to
the Brownian motion but is related to the wave function of the
metastable Cooper pairs, the Ginzburg-Landau order parameter
$\Psi(\mathbf{r},t)$ which for the normal phase is a time and
space-vector dependent stochastic function. When the temperature
is below the critical one $T<T_c$ both the amplitude of the wave
function and the number of fluctuation Cooper pairs increase with
the time. Such wave amplification is analogous to the lasing
process in lasers or to the dynamics of the Bose condensation.
This is the precursor of the transition of the superconductor in
the superconducting phase which has infinite conductivity; the
electric current can flow without an external voltage in the
superconducting phase. However, the applied DC electric field
prevents fluctuation Cooper pairs to condense in a coherent
superconducting phase; superconducting phase cannot exist under
external voltage. Once born, the electric field accelerates the
fluctuation Cooper pairs and their kinetic energy increases.
However, the decay rate of the Cooper pairs is energy dependent
and increases with the energy. Like on a highway big velocity
increases the probability of accidents. Roughly speaking, the life
of faster Cooper pairs is shorter and the electric field finally
destroys the accelerated Cooper pairs. During their life
metastable Cooper pairs carry significant electric current
(fluctuation current) comparable and even bigger than the current
of normal charge carriers electrons or holes (normal current); the
total current is the sum of fluctuation and normal current.

We consider as very instructive to give also popular explanations
using the analogy with ideas from two atmospheric phenomena which
have no direct link in real meteorology: rains and meteors; as we
stressed it will be only analogy. (1) Imagine a very humid
atmosphere supercooled below the condensation temperature. Small
droplets continuously grow in size and it begins raining. We have
different droplets with different sizes and velocities that create
the current density of the rain -- a total amount of water passing
through a square meter per one second. (2) Second phenomenon is
related to meteorites falling from the cosmic space; moving with
high velocities stones burn in the atmosphere due to the heating
by the friction force. Imagine now that Earth acceleration is many
orders of magnitude higher (for the real meteorology it is
unrealistic) that makes big rain droplets moving with high
velocity evaporate by heating from the friction force in bigger
extent. This sets limits to the maximal size for falling droplets
and the rain could be a continuous process with a current density
determined by the acceleration force. The subtle question here is
what will happen if acceleration will slightly decrease. The
smaller acceleration will lead to a bigger size of the largest
droplets and as a consequence the debit of the rain will increase.
This is an example of how NDC appears -- small decrease in the
driving force cause increasing of the current. The analog of the
droplets is the square of modulus of the Fourier components of the
effective Ginzburg-Landau function $|\Psi_\mathbf{k}(t)|^2.$ When
a superconductor is cooled below the critical temperature $T_c$ in
the bulk of the superconductor starts a process similar to Bose
condensation. Electrons start to condensate in Cooper pairs and
many Cooper pairs can have a common momentum $\hbar \mathbf{k}$ --
this is the analog of a rain droplet having a velocity
proportional to its momentum.

After this analogy and a qualitative description of the main
processes we can analyze in detail the appearance of the NDC --
the current increasing when the electric field decreases. Such a
behavior is opposite to the ohmic conductivity.

When the electric field is smaller, the acceleration is slower,
the decay rate of fluctuation Cooper pairs is slower, and as a
result the volume density of the fluctuation Cooper pair is higher
and the current is higher too. The acceleration, creating the
decay of Cooper pairs, is an analog of the evaporation of droplets
by the heating force from our above-mentioned example. So, the
electric field accelerates and destroys Cooper pairs and the
superconductor cannot move to superconducting state. Thus it is
logically that the electric current density $j(E)$ increases when
the electric field $E$ decreases, i.e. the differential
conductivity of the material is negative
\begin{equation}
\varsigma_\mathrm{diff}(E)=\frac{dj(E)}{dE}.
\end{equation}

In such a non equilibrium situation supercooled superconductor
will have NDC due to the electric field dependence of the density
of metastable Cooper pairs.

Here, by way of illustration we will consider another analogy of
the creation of electric oscillations by supercooled
superconductors. In some sense, the active part of a
superconductor with NDC operates as a laser. However, there is a
need to pump the energy in the active medium in order to make
lasers operating. Now, let us consider how the superconductor is
lasing. Below $T_c$, the ground state of the material is the
superconducting state having the lowest free energy and the normal
state (nonsuperconducting state with an Ohmic resistivity) is the
``excited state'' having higher energy and generating electric
oscillations. The energy for these oscillations comes from the
constant electric field but in some sense the excited state is
reached by cooling. For high-$T_c$ superconductors the cooling can
be done by liquid nitrogen that is substantially cheaper. The
initial demonstrations of the effect of the NDC, however, can be
realized at helium temperatures by technologically more convenient
conventional superconductors.

Like the current, the total differential conductivity is a sum of
its fluctuation and normal parts. The normal part of the
differential conductivity $\varsigma_N(T)$ weakly depends on the
electric field. In the next subsection we will describe the
state-of-the-art theory of the differential conductivity of the
fluctuation Cooper pairs.

\subsection{Formulas for the differential conductivity}
%
For a comprehensive contemporary review on the properties of
superconductors see the book edited by Benneman and Ketterson
which starts with a review article by Larkin and Varlamov on the
fluctuation phenomena in superconductors.\cite{Larkin02} There can
be found detailed explanations of the Bardeen, Cooper and
Schrieffer (BCS) theory of superconductivity and also of the time
dependent Ginzburg-Landau (TDGL) theory for the order parameter of
superconductors. Ref.~\onlinecite{Mishonov00} is another review
especially devoted to the Gaussian fluctuation in superconductors.
The Boltzmann equation for the fluctuation Cooper pairs was
derived\cite{Mishonov:96} in the framework of TDGL theory. For the
case of strong electric fields Boltzmann equation was solved in
Ref.~\onlinecite{Mishonov:02}, and a general formula for the
fluctuation current was derived as well; see also the references
therein. The same result was rederived\cite{Mishonov:03} directly
from the TDGL theory. Our formula for small electric fields below
$T_c$ is similar to the formula by Go'kov (1970) which was
however, directly derived\cite{Gor'kov:70} within the framework of
BCS theory; see also the work by Tucker and
Halperin.\cite{Tucker:71} Differentiating the formula for the
fluctuation current\cite{Mishonov:02,Mishonov:03}
\begin{equation}
j_\mathrm{fl}(E_x)
=\frac{e^2\tau_\mathrm{rel}E_x}
      {16\hbar\left[\pi^{1/2}\xi(0)\right]^{D-2}}
\int_0^\infty
\frac{\exp\left(-\epsilon u-gu^3\right)}
     {u^{(D-2)/2}}
\,d u,
\end{equation}
we obtain the formula for the total differential conductivity
\begin{eqnarray}
\label{SigmaDiff}
\varsigma_\mathrm{diff}(E)
&=&\varsigma_N(T)
 +\frac{e^2\tau_\mathrm{rel}}{16\hbar\left[\pi^{1/2}\xi(0)\right]^{D-2}}\\
 &&\quad\times\int_0^\infty \frac{\exp\left(-\epsilon
  u-gu^3\right)}{u^{(D-2)/2}}\, \left(1-2gu^3\right)d u,\nonumber
\end{eqnarray}
where $D$ is the dimension of the space, $e$ is the electron
charge, $\xi(0)$ is the Ginzburg-Landau (GL) coherence length of
the superconductor, $\tau_\mathrm{rel}$ is a dimensionless
constant which describes how long the fluctuation Cooper pairs
live in comparison with the prediction of BCS theory, $k_B$ is the
Boltzmann constant, $U$ is the voltage difference, $L$ is the
length of the sample, and
$$
\epsilon=\frac{T-T_c}{T_c},
\quad g=\frac{f^2}{12},
\quad f=\frac{\pi}{8}\,\frac{eE\xi(0)}{k_BT_c}\,\tau_\mathrm{rel},
\quad E=\frac{U}{L}.
$$
The analysis of Eq.~(\ref{SigmaDiff}) shows that below $T_c$ where
$\epsilon<0$ the differential conductivity is really negative as
it is necessary for the work of the suggested current oscillator.
The GL theory is formally applicable only close to $T_c$ for
$|\epsilon|\ll 1$, but qualitatively its results can be used even
far from $T_c.$ In other words, the differential conductivity will
remain negative even if the accuracy of the TDGL formula
Eq.~(\ref{SigmaDiff}) is not very high. We have to mention that
TDGL equation is derived from BCS theory as a result of some
approximations and for some cases it could be only a convenient
model equation. We also wish to point out that the dimension of
the current density depends on the dimension of the space
$[j_D]=\mathrm{A/m}^{D-1}$: for a bulk sample
$[j_3]=\mathrm{A/m^2},$ for thin films with a thickness
$d_\mathrm{film}\ll \xi(\epsilon)$ $[j_2]=\mathrm{A/m},$ and for a
wire with a cross-section $\ll\xi^2(\epsilon)$ the current density
is just the current $[j_1]=\mathrm{A}.$ Here
\begin{equation}
\xi(\epsilon)=\frac{\xi(0))}{\sqrt{|\epsilon|}}
\end{equation}
is the temperature dependent coherence length. It is also
convenient to introduce a temperature dependent Cooper pair
life-time $\tau(\epsilon)$
\begin{equation}
\tau(\epsilon)=\frac{\tau(0)}{|\epsilon|}, \qquad
\tau(0)=\frac{\pi}{16}\frac{\hbar}{k_BT_c}\tau_\mathrm{rel},
\end{equation}
where the numerical coefficient $\pi/16$ is a result of the
microscopic BCS theory. Analogously, it is convenient to introduce
a dimensionless temperature dependent electric field
\begin{equation}
f_\epsilon=\frac{f}{|\epsilon|^{3/2}}
=|e^*E|\xi(\epsilon)\tau(\epsilon)/\hbar\ll1,
\end{equation}
where $|e^*|=2|e|$ is the charge of the Cooper pair. The dummy
parameter of the integration in Eq.~(\ref{SigmaDiff}) has a
physical meaning of a dimensionless time $u=t/\tau(0),$ and
analogously, one can introduce another dimensionless time
$v=t/\tau(\epsilon).$

The present theory is applicable for every superconductor which is
homogeneous enough in order to avoid nucleation of the
superconducting phase. However, we consider as the most promising
the cuprate high-$T_c$ superconductors containing as a main
structural detail superconducting CuO$_2$ planes, such as
YBa$_2$Cu$_3$O$_{7_\delta}$ and Bi$_2$Sr$_2$CaCu$_2$O$_8$
superconductors which have $T_c\approx 90$~K and can be cooled by
liquid nitrogen using working temperatures $T=80$~K and reduced
temperature $\epsilon\simeq -0.1$. The coherence lengths in
CuO$_2$ plane are typical for other 90~K cuprates
$\xi_{ab}(0)\simeq 2$~nm. All high-$T_c$ cuprates have a
significant anisotropy but Bi$_2$Sr$_2$CaCu$_2$O$_8$ is extremely
anisotropic. Even for small reduced temperatures $|\epsilon|\simeq
0.1$ the coherence length perpendicular to the CuO$_2$ plane can
be smaller than the distance $s$ between double planes CuO$_2.$ In
this case, every double planes operate approximately as an
independent two dimensional (2D) layer and the number of layers
$N_\mathrm{l}$ depends on the film thickness
$N_\mathrm{l}=d_\mathrm{film}/s.$ If the superconductor is a strip
with a width $w$, patterned from a layered superconductor we have
for the total current
\begin{equation}
I=\frac{wd_\mathrm{film}}{s}j_2(E).
\end{equation}
Such a way for the differential conductance of the sample we
obtain
\begin{equation}
\label{SampleConductivity}
\sigma_\mathrm{diff}(U)
=\frac{di(U)}{dU}=
\sigma_N+\frac{e^2\tau_\mathrm{rel}wd_\mathrm{film}}
              {16\hbar s |\epsilon|}
S(g_\epsilon),
\end{equation}
where the universal function
\begin{equation}
\label{UniversalFunction}
S_\mathrm{diff}(g_\epsilon)
=\int_0^\infty \left(1-2g_\epsilon v^3\right)
\exp\left(\mathrm{sign}(-\epsilon)-g_\epsilon v^3\right) d v
\end{equation}
have to be calculated only once for the
$\mathrm{sign}(-\epsilon)=\pm 1.$ The negative differential
conductivity arises only for supercooled below $T_c$
superconductors, and in this case $\mathrm{sign}(-\epsilon)=1.$ In
Eq.~(\ref{UniversalFunction}) the electric field is parameterized
by the dimension parameter
\begin{equation}
g_\epsilon=\frac{g}{|\epsilon|^3}
=\frac{1}{12}f_\epsilon^2
=\frac{1}{12|\epsilon|^{3}}
\left(\frac{\pi eU\xi(0)\tau_\mathrm{rel}}{8k_BT_cL}\right)^2.
\end{equation}
In order to have a significant negative differential conductivity
this parameter should be small enough $g_\epsilon\ll 1.$ This
means that close to the critical region the applied DC voltage $U$
should be small enough.

The fluctuations are stronger in low dimensional systems, that is
why another realization of the negative differential conductivity
in superconductors could be a nanostructured stripe of
conventional superconductor on nanowires. Consequently, we have to
use one dimensional (1D) formula for the current. Analogously, for
layered superconductors we can use Lawrence-Doniach theory which
can be interpolated by some space dimension $2\le D\le3.$ It is
necessary to use disordered conventional superconductors or
cuprates in order to have smaller normal current.

\section{Description of the oscillations}
%
In order to illustrate how electric oscillations can be generated
by a supercooled superconductor we will use the simplest possible
electric scheme used in generators with tunnel
diodes.\cite{Chow:64} For a pedagogical explanation of this scheme
and the Van der Pol equation see also the excellent
textbook.\cite{Enns:01}

The superconductor is connected in parallel with one resistor
having resistance $R$ and with one capacitor having capacity $C$.
Those 3 elements are sequentially connected in a circuit with one
inductance $L$ and a battery with electromotive force
$\mathcal{E}.$ For a static current the voltage on the
superconductor, the capacitor and the resistor is just the voltage
of the battery $U=\mathcal{E}.$ In this static case, the voltage
of the inductance is zero.

Let us now consider what will happen if the superconductor is
supercooled. The whole circuit is cooled below $T_c,$ but we
consider that in the beginning superconductor is in the normal
state. Imagine that it is heated by a short current or a laser
impulse. We will analyze the fluctuations of the voltage of the
superconductor taking into account the static solution
$U(t)=\mathcal{E}+x(t).$ The deviation from the static solution
$x\equiv U(t)-\mathcal{E}$ obeys the differential equation
\begin{equation}
C\frac{d^2}{dt^2}x
+\left[\frac{1}{R}+\sigma_N
+\sigma_\mathrm{diff}(\mathcal{E}+x)\right]\frac{d}{dt}x
+\frac{1}{L}x=0,
\end{equation}
cf. Ref.~\onlinecite{Enns:01}. After introducing the auxiliary
variable
\begin{equation}
y(t)\equiv\frac{d}{dt}x(t)
\end{equation}
the second equation of this system of ordinary differential
equations reads as
\begin{equation}
\frac{d}{dt}y(t)=-\nu(v)y-\omega^2 x,
\end{equation}
where
\begin{equation}
\nu(x)\equiv \frac{1}{C}
\left[\frac{1}{R}+\sigma_N
 +\sigma_\mathrm{diff}(\mathcal{E}+x)\right],
\quad \omega=\frac{1}{\sqrt{LC}}.
\end{equation}
For a moderate accuracy necessary for the modelling of electronic
circuits we can use some adaptive Runge-Kutta
method,\cite{Press:01} or a simple empirical formula for the time
step which follows the characteristic frequencies of the circuit
\begin{equation}
\Delta t= 0.1/\sqrt{\nu^2(x)+4\omega^2}.
\end{equation}
The physical restrictions for the high frequencies are related
only to applicability of the TDGL equation and the static formulae
for the current response. The static response approximation used
for the derivation of Eq.~(\ref{SigmaDiff}) is applicable for
$\omega\tau(\epsilon)\ll1$. This means that for a high-$T_c$
superconductor this generator can operate for the whole radio
frequency range and even in the far infrared region
\begin{equation}
\omega\ll |\epsilon| k_BT_c/\hbar\simeq 1/\tau(\epsilon).
\end{equation}
Let us now describe the appearance of the oscillations when the
superconductor is supercooled. The total differential conductivity
of the circuit becomes zero at the temperature $T_g$, determined
by solution of the equation
\begin{equation}
-\sigma_\mathrm{diff}(\mathcal{E},T_g)=\frac{1}{R}+\sigma_N(T_g).
\end{equation}
Further cooling leads to appearance of NDC, the static solution
$x(t)=0$ looses stability, and the voltage in the circuit starts
to oscillate. The main difference with the tunnel diode devices is
that for superconductors we have no definite region of negative
conductivity as a function of the voltage. The current
continuously increases when the voltage decreases down to zero
voltage. It is quite possible that amplitude of the oscillations
will be limited only by ohmic heating of the sample. For thin
films the heat current is determined mainly by the boundary
resistance $\mathcal{R}_h$ of the interface of the superconductor
and the insulator substrate. For high frequencies we can average
the dissipated power and calculate the local increasing of the
temperature of the superconductor above the ambient temperature
\begin{equation}
\Delta T = \frac{\langle I(t)U(t) \rangle_t}{\mathcal{R}_h}.
\end{equation}
Such a way we obtain a self-consistent equation for the reduced
temperature
\begin{equation}
\epsilon=\frac{\Delta T + T - T_c}{T_c}
\end{equation}
which have to be substituted in the formula for the differential
conductivity Eq.~(\ref{SampleConductivity}). The complex problem
of the temperature and electric field oscillations can be easily
simulated on a computer in order to optimize the parameters of the
device and the initial stage of the experimental research.

One can also speculate what will happen if we start with a
superconducting sample. Applied voltage will destroy the
superconductivity but it is also possible that a space
inhomogeneous state will appear. It is difficult to predict the
behaviour of the system when having a problem related to the
domain structure. That is why an experimental investigation is
needed.

\section{Performance of the generator}
The most important prerequisite for the realization of NDC by
supercooled superconductors is to keep the superconductor in the
metastable normal state by preventing its active part from
transition to the superconducting state which is thermodynamically
stable below $T_c$. The sample has to be clean from defects, for
instance pin holes, that can nucleate locally the
superconductivity. Special efforts have to be applied to the
contacts of the superconductor sample where the current density is
low and these regions could be a source of nucleation of
superconducting domains. We consider that only the central working
region of the sample should be superconducting. The contact area
should have inserted depairing defects. Such defects could be the
magnetic impurities for conventional superconductors, Zn in
CuO$_2$ plane, etc. Oxygenation of YBa$_2$Cu$_3$O$_{7-\delta}$
superconductors or changing stoihiometry are also a tool to change
$T_c$ locally. Overdoped and underdoped cuprate thin films will
have opposite behaviour with respect to the stability of
appearance of space inhomogeneous domains of the superconducting
and normal phases.

The amplitude of the oscillations can also be restricted by a
current limiter with a maximal current $I_c(T)$, a narrow
superconducting wire, Josephson junction, or narrow
superconducting strip sequentially switched to the inductance of
the resonance circuit. In a rough approximation the resistance of
the limiter $R_\mathrm{i}=R_0\theta(I_c(T)-I)$ is switched when
the current $I(t)$ becomes bigger than the critical $I_c(T)$. Such
additional amplitude dependent dissipation will prevent the sample
to pass into normal state when $U(t)=\mathcal{E}+x(T)=0.$ For such
a small modification of the circuit we have to solve the system of
equations
\begin{eqnarray}
\mathcal{E}&=& L\frac{d I}{dt}+R_0\,\theta(I_c(T)-I(t))\,I(t)+U(t),\\
I(t)&=&\left(\frac{1}{R}+\frac{1}{R_N}\right)U(t)
+C\,\frac{d U(t)}{d t}
+I_\mathrm{fl}[U(t)],\nonumber
\label{to be added}
\end{eqnarray}
where, if necessary, the Boltzmann equation can be solved in the
general case in order to obtain the high frequency functional for
the fluctuation current $I_\mathrm{fl}[U(t)]$ and eventually the
self-interaction between fluctuation Cooper pairs has to be taken
into account.

In the paper by Gor'kov\cite{Gor'kov:70} were mentioned some early
experiments for observation of oscillations in superconductors
close to $T_c$. Those experiments give a hint that creation of
oscillations in supercooled superconductor is also possible. But
the problem requires detailed experimental investigation. Very
often NDC leads to space inhomogeneities instead of time
oscillations and we consider that both regimes can be realized. To
avoid the appearance of space inhomogeneities one can also use
narrow (nanostructured) region of supercooled superconductor which
will be analogous to the tunnelling region of a tunnel diode. The
length of the superconducting area should be comparable with the
coherence length of the superconductor $\xi(\epsilon).$

Fluctuation effects in conventional superconductors are very weak,
with the exception of some very disordered films with a high Ohmic
resistance. They cannot be effectively used in THz region due to
their relatively smaller critical temperature $T_c$, the
superconducting gap and the energy scale in general. However,
systematic investigations of conventional superconducting
nanostructures can be very important step for creation of electric
oscillators with a superconductor as an active element. In
experiments with gaseous plasma and semiconductors the NDC can be
reached by various viable methods and it is not strange that the
same could happen for superconductors. The investigation of
current-voltage characteristics of superconducting
nanowires\cite{Michotte:03} is promising. We consider that for
superconducting nanowires could be taken into account the phase
slip centers, the domains of different temperature, or even some
possible modification of time-dependent Ginzburg-Landau theory
might be done. A detailed physical picture can be drawn only after
detailed investigations. We suggest conventional superconductors
to be used to demonstrate that NDC can be reached and thus to
stimulate further investigation of cuprate nanostructures.
Successful experiments with high-$T_c$ superconductors can trigger
significant applications in the THz range as mentioned above. When
the current oscillator is realized by a cuprate the resonance
circuit (the capacitor, the inductive coil and the current
limiter) can be patterned in a single chip from the same submicron
cuprate layer. We can generate electric oscillations and NDC to be
well inside the THz region only by nanostructured cuprates because
the maximal frequency at which NDC can exist is proportional to
the maximal value of the superconducting gap and critical
temperature $T_c.$

\section{Possible applications}
%
The described principle of generation of current oscillations can
be realized in superconducting electronics. In general, the
negative differential conductivity (NDC) could be a useful tool in
the electronics using superconductors as an active element.

We consider that the generation of submillimeter electromagnetic
waves by high-$T_c$ superconductors is quite possible. Thin
superconducting layers can be electrostatically doped and in this
context some preliminary research on superconducting field effect
transistors can be extremely helpful for modulation of the
oscillations. There are good prospectives for application in the
wireless communications. We consider also that in the regime of
supercooling some old samples could become working transistors.

Fluctuations are more important for the low dimensional systems.
The fact that NDC can be observed in conventional nanostructured
superconductors is promising. In this case the 1D theory can be
directly used or one can easily perform summation on the
perpendicular modes of the Cooper pair wave guide. Shortly said,
the combination of nanostrip, capacitor and inductance patterned
into one chip or the plasma modes of the superconducting stripe,
could be considered as the smallest current oscillator performed
for demonstration purposes.

The device constructed upon our instructions will be very
sensitive to the temperature and might be stabilized to oscillate
near the critical temperature $T_c$. It should not only prove our
theory but could be used as well as a superconducting bolometer.
The sensitivity to temperature variations around $T_c$ opens
opportunities for optoelectronic applications.

The negative differential conductivity can be investigated in the
case of zero temperatures $T\ll T_c$ and the critical behaviour in
small electric field is appropriate for investigation of the
quantum criticality. This is an evidence that development of the
applied research can in return stimulate further development of
the initializing theory.

The described idea can be realized not only for the in-plane
conductivity of MgB$_2$ and CuO$_2$ containing high-$T_c$
superconductors but also for the currents in the so called
c-direction perpendicular to the CuO$_2$ layers. For
Bi$_2$Sr$_2$CaCu$_2$O$_8$ the NDC can be coupled with plasma
resonances having frequencies lower than the superconducting gap.
This coupling leads to natural realization of voltage induced FIR
oscillations with low dissipation.

The nonlinear conductivity in supercooled regime can be used also
for frequency mixers and transistors. The possibility to operate
in terahertz range (loosely defined by the frequency range of 0.1
to 10 THz) using hight-$T_c$ superconductors looks very promising.
Recently Ferguson and Zhang\cite{Ferguson:02} mentioned that the
lack of high-power, low-cost, portable room-temperature THz source
is the most significant limitation of the modern THz systems. They
consider that the narrow band THz sources are crucial for
high-resolution spectroscopy applications. In addition, the
authors are stressing that this kind of sources have broad
potential applications in telecommunications and are particularly
attractive for extremely high bandwidth intersatellite links. Our
invention is designed to fill this gap with its simple theoretical
explanation and easy for performance practical proposal.

The THz region of the electromagnetic waves (the frequencies
between 100 GHz and 10 THz or wavelengths between 3 mm and 30
$\mu$m lies between radiofrequencies and optics and up to now is
not so well developed for applications and research as
neighbouring frequency ranges. The state of the art review on the
application and sources of THz radiation and the need of
improvement and new sources is given in a recent review article by
Mueller.\cite{Mueller:03} The author points out that practical
application of THz radiation is in initial stage and only some
reliable sources are available. That is why every new principle of
THz wave generation can be not only significant theoretical input
but will have important technical applications in the near future.
In the review\cite{Mueller:03} is mentioned also that recently the
investigators are pursuing potential THz-wavelength applications
in many fields: quality control; biomedical imaging; THz
tomographic imaging in mammography; passengers' screening for
explosives at the airports; detecting the presence of cancerous
cells; complex dynamics involved in condensed matter physics;
molecular recognition and protein folding; environmental
monitoring; plasma diagnostics; Antarctic submillimeter telescope
which will be used to measure interstellar singly ionized nitrogen
and carbon monoxide during the polar winter; significant part of
the photons emitted since the Big Bang fall also in THz region -
continuous-wave THz sources can be used to help study these
photons;  THz imaging using time domain spectroscopy developed in
Lucent Technologies' Bell Laboratories - it uses the greatly
varying absorption characteristics from material to material;
NASA's AURA satellite measuring the concentration and distribution
of hydroxyl radical (OH-) in the stratosphere, a crucial component
in the ozone cycle, etc.

At the end, we would like to draw reader's attention to another
possible application. In a semiconductor nanostructure with two
dimensional electron gas and an appropriate grating coupler THz
electromagnetic waves can be transmitted in hyper sound phonons --
such a way we will have an ersatz phonon laser useful for phonon
spectroscopy as well.\cite{Mishonov:90}

In conclusion we consider that investigation of voltage biased
conductivity of nanostructured superconductors is very perspective
theme of the fundamental science promising viable variety of
technical applications.

\section{Acknowledgments}
We would like to especially thank V.~Mishonova who has initiated
this challenging project and contributed to its completion.
Special thanks go to W.~Bird and I.~Roelants for their cooperation
and stimulating discussions. One of the authors (TMM) is thankful
to D.~Cocke for pointing out the Ref.~\onlinecite{Mueller:03} and
to T.~Donchev for the critical reading of the manuscript.

\end{document}